\documentclass{article}

\usepackage{PRIMEarxiv}
\usepackage[utf8]{inputenc}
\usepackage[T1]{fontenc}
\usepackage{amsmath,amssymb,amsfonts,amsthm}
\usepackage{booktabs,multirow,tabularx,array}
\usepackage{graphicx}
\usepackage{xcolor,xspace}
\usepackage{enumitem}
\usepackage{microtype}
\usepackage[numbers]{natbib}
\usepackage{placeins}
\usepackage{hyperref}
\usepackage{url}

\newcommand{\shorttitle}{Deep Cross-Division Network for Industrial Ranking}
\hypersetup{
  hidelinks,
  pdftitle={Deep Cross-Division Network for Product-and-Ratio Feature Interactions in Industrial Ranking},
  pdfauthor={Mikhail Trapeznikov, Sofia Gulevskaia, Maksim Utushkin, Aleksandr Poslavsky, Alexander D'yakonov},
  pdfsubject={Recommender systems; learning to rank},
  pdfkeywords={recommender systems, learning to rank, feature interactions, serving efficiency, Deep and Cross Network}
}
\newtheorem{theorem}{Theorem}
\providecommand{\Description}[2][]{}

\newcolumntype{Y}{>{\raggedright\arraybackslash}X}
\newcommand{\hiformer}{\textsc{Hiformer}\xspace}

\begin{document}

\title{Deep Cross-Division Network for Product-and-Ratio Feature Interactions in Industrial Ranking%
  \thanks{Accepted at the Twentieth ACM International Conference on Web Search and Data Mining (WSDM '27), February 15--19, 2027, Hong Kong. This is a preprint version. The proceedings version has DOI: \href{https://doi.org/10.1145/3862322.3863643}{10.1145/3862322.3863643}.}}

\author{
  Mikhail Trapeznikov\\
  AI VK\\
  Lomonosov Moscow State University\\
  Moscow, Russia\\
  \texttt{m.trapeznikov@vk.team}\\
  \And
  Sofia Gulevskaia\\
  AI VK\\
  Lomonosov Moscow State University\\
  Moscow, Russia\\
  \texttt{s.gulevskaia@vkteam.ru}\\
  \AND
  Maksim Utushkin\\
  AI VK\\
  Moscow, Russia\\
  \texttt{mak.utushkin@gmail.com}\\
  \And
  Aleksandr Poslavsky\\
  AI VK\\
  Moscow, Russia\\
  \texttt{dr.slink@vk.com}\\
  \And
  Alexander D'yakonov\\
  AI VK\\
  Moscow, Russia\\
  \texttt{djakonov@mail.ru}\\
}

\maketitle

\begin{abstract}
Learning effective feature interactions is essential for industrial recommender systems. DCN and DCNv2 efficiently automate polynomial feature crossing. However, their cross layers construct interactions using only non-negative integer powers of the input features, with the maximum degree controlled by network depth. Consequently, they cannot directly represent the reciprocal terms required for ratios, rates, and normalized counts
unless such features are explicitly provided as inputs.
We introduce DCDN, a Deep Cross-Division Network that augments a DCNv2-style cross layer with a division branch. DCDN jointly learns product, ratio, and mixed product–ratio interactions.
We characterize the representable interactions as multivariate Laurent polynomials and show that a $K$-layer DCDN generates Laurent monomials of $\ell_1$-order up to $K+1$, with signed degree between $-K$ and $K+1$.
We evaluate DCDN against a production-optimized neural-ranking baseline built around DCNv2 and supplied with manually engineered ratio features. Replacing the DCNv2 block with DCDN allowed us to eliminate the entire class of manually engineered ratio features from the model input, reducing the number of input features by $40.1\%$. DCDN is deployed in the production ranking stack of a major short-video service. In a live A/B test on this service, DCDN increased likes by $3.25\%$, shares by $16.26\%$, total view time (TVT) by $0.33\%$, and clicks to open fullscreen by $0.38\%$; all changes were statistically significant.
The number of shown videos
decreased slightly ($-0.24\%$).
These results demonstrate that DCDN can improve user engagement while eliminating reliance on manually engineered ratio features. Our code is publicly released at \href{https://github.com/rw404/DCDN}{https://github.com/rw404/DCDN}.
\end{abstract}


\section{Introduction}
Large-scale ranking systems rarely optimize predictive quality in isolation. A ranker scores many candidates under a strict latency budget and often at very high request volume. A model that improves an offline metric can therefore be unattractive in production if it reduces throughput, increases tail latency, or requires substantially more resources. In our production stack, model application accounts for approximately $20\%$ of the end-to-end recommendation response time. Model selection is consequently a quality--latency decision rather than an unconstrained accuracy contest.

Part of this trade-off comes from the feature set itself. Production rankers commonly augment user, item, and context signals with derived statistics that encode domain knowledge. Ratio features are a familiar example: click-through rate normalizes clicks by impressions, and analogous features normalize other engagement counts by exposure or opportunity. Such features can be highly predictive because they expose a useful relation explicitly, but they also widen the input and increase the work done by downstream interaction layers. We distinguish a base feature set $B$ from a set $R$ of engineered ratio features whose numerator and denominator operands remain available in $B$. Removing $R$ makes the representation cheaper, but can discard useful signal.

We study whether some of that signal can instead be recovered through the interaction operator. DCDN extends the matrix CrossNet layer of DCNv2~\cite{wang2021dcnv2} with a reciprocal crossing anchored at the original input. The additional branch makes relations between retained operands easy to express while preserving the surrounding ranking architecture. The design is intentionally specialized: it targets feature regimes in which ratio-like interactions matter.

AFN~\cite{cheng2020afn} and EulerNet~\cite{tian2023eulernet} already support negative or non-integer orders through more general parameterizations. The practical question is whether a direct reciprocal bias can provide a better cost--quality trade-off for ratio-like interactions in ranking data.

We examine that question from four angles. A controlled quotient task first tests what the reciprocal parameterization actually learns. We then evaluate DCDN on Yambda, KuaiRec, KuaiRand, and VK-LSVD against a broad set of feature-interaction models using task-appropriate ranking metrics. Separate latency and capacity-matched comparisons distinguish accuracy, model size, and measured inference cost. Finally, we replace the DCNv2 block in a production multi-target short-video ranker and evaluate both online user metrics and serving performance.

Our contributions are:
\begin{itemize}[leftmargin=*,nosep]
  \item We introduce a reciprocal CrossNet interaction and characterize the exact Laurent-monomial support of a $K$-layer DCDN stack, including tight signed-degree bounds from $-K$ to $K+1$.
  \item We evaluate DCDN on a controlled quotient task and four public recommendation tasks, comparing it with DCNv2 and representative interaction models while separating full-feature-view, capacity-matched, and measured-latency comparisons.
  \item We deploy DCDN in a large-scale short-video ranker, substantially reducing engineered ratio features while improving both user-facing and serving metrics.
\end{itemize}

\FloatBarrier
\section{Related Work}
\emph{Feature interaction modeling.}
Feature interaction modeling is central to CTR prediction and ranking. Factorization Machines model pairwise interactions with shared latent factors~\cite{rendle2010fm}; DeepFM combines factorization with a deep component~\cite{guo2017deepfm}; xDeepFM constructs bounded high-order interactions with a Compressed Interaction Network~\cite{lian2018xdeepfm}; and AutoInt learns interactions with self-attention~\cite{song2019autoint}. Wukong scales explicit interaction capacity through stacked factorization-style blocks~\cite{zhang2024wukong}. These models span a wide range of interaction families and serving costs, which makes them useful references for evaluating whether a more specialized operator can occupy a competitive quality--efficiency regime.

\emph{Deep \& Cross Networks.}
DCN~\cite{wang2017dcn} and DCNv2~\cite{wang2021dcnv2} are the direct architectural predecessors of DCDN. Their cross layers repeatedly combine a learned projection of the current representation with the original input, yielding bounded-order multiplicative interactions in a form that has proved practical at web scale. Later work has modified this family in several directions: FCN increases the growth rate of explicit high-order interactions~\cite{li2026fcn}, DCN$^2$ studies explicit cross layers together with implicit collision effects~\cite{skrlj2025dcn2}, and recent efficiency-oriented variants reduce or restructure the cross computation~\cite{tkach2026context,li2026mldcn}. DCDN takes a different route: it keeps the CrossNet recursion but adds a reciprocal anchor, so normalized relations can be formed directly from retained operands.

\emph{Adaptive-order and arithmetic interactions.}
AFN learns adaptive feature orders after a logarithmic transformation, allowing negative and fractional exponents~\cite{cheng2020afn}; AFN+ combines this component with a deep predictor. EulerNet uses Euler's formula to support real-valued interaction orders without the same non-negativity restriction on the original representation~\cite{tian2023eulernet}. These methods establish that negative powers and division-like dependencies are already within the scope of neural feature-interaction models. They also provide an important contrast to DCDN: AFN and EulerNet are substantially more general, whereas DCDN deliberately hard-codes only the reciprocal direction. This distinction motivates our controlled quotient experiment: beyond asking whether a ratio is representable, we test whether the corresponding dependency is easy to learn from its operands.

\emph{Feature selection and efficient ranking.}
A complementary line of work reduces serving cost by changing which features are retained or how they are processed. AdaFS adaptively emphasizes predictive features~\cite{lin2022adafs}, while OptFS jointly considers feature selection and interaction modeling to remove unnecessary computation~\cite{lyu2023optfs}. Rank-aware cross-network decomposition targets the cost of the interaction operator itself~\cite{tkach2026context}. Our setting connects these two concerns: removing engineered ratios reduces input and interaction cost, but may lower ranking quality; reciprocal crossing attempts to recover part of that signal from the operands that remain, shifting the burden from the feature pipeline to the model's inductive bias.

\FloatBarrier
\section{DCDN: Reciprocal Cross Interactions}
\subsection{Ratio-feature setting}
Let $B=\{b_1,\ldots,b_p\}$ denote the base features available to the ranker and $R=\{r_1,\ldots,r_m\}$ a set of engineered ratios computed from operands already contained in $B$. A typical ratio is $ r_j=\frac{b_l}{b_k}.$
We distinguish the full and reduced feature views
\begin{equation}
  F_{\mathrm{full}}=B\cup R,
  \qquad
  F_{\mathrm{reduced}}=B.
  \label{eq:feature-sets}
\end{equation}
The reduced view removes only the derived ratio itself; its numerator and denominator remain available. DCDN is designed to make such normalized relations easy to reconstruct without requiring every useful ratio to be supplied as an explicit input feature.

\subsection{From CrossNet to reciprocal crossing}
\begin{figure}[htbp]
    \centering
    \includegraphics[width=0.72\linewidth]{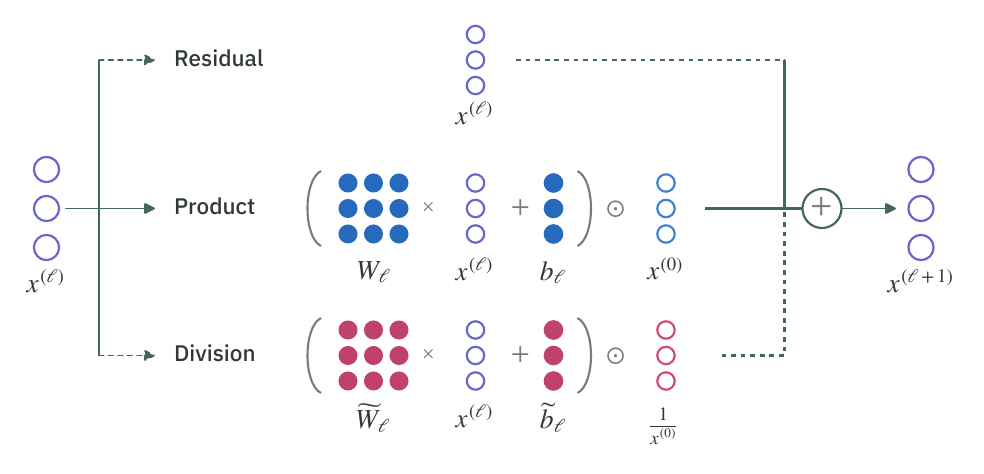}
    \caption{
        The DCDN cross operator. The reciprocal term augments
        the multiplicative crossing within the cross block.
    }
    \Description{
        Diagram of a DCDN cross layer with a standard multiplicative
        branch anchored at the original input and a reciprocal branch
        anchored at its element-wise reciprocal.
    }
    \label{fig:dcdn-operator}
\end{figure}

For a matrix CrossNet layer, let $x^{(0)}\in\mathbb{R}^d$ be the original input representation and $x^{(\ell)}$ the representation at layer $\ell$. The standard update is
\begin{equation}
  x^{(\ell+1)}
  =x^{(\ell)}+
  \bigl(W_{\ell}x^{(\ell)}+b_{\ell}\bigr)\odot x^{(0)}.
  \label{eq:dcn-layer}
\end{equation}
DCDN keeps this multiplicative cross and adds a second branch anchored at the element-wise reciprocal of the original input:
\begin{equation}
  \begin{aligned}
  x^{(\ell+1)} ={}& x^{(\ell)}
  +\bigl(W_{\ell}x^{(\ell)}+b_{\ell}\bigr)\odot x^{(0)}
  +\bigl(\widetilde W_{\ell}x^{(\ell)}+\widetilde b_{\ell}\bigr)
     \odot \frac{\mathbf{1}}{x^{(0)}},
  \end{aligned}
  \label{eq:dcdn-layer}
\end{equation}
where division in the reciprocal term is element-wise, $W_{\ell}, \widetilde{W}_{\ell} \in \mathbb{R}^{d \times d}$,
and $b_{\ell}, \widetilde{b}_{\ell} \in \mathbb{R}^{d}$. The reciprocal term changes only the cross operator; it does not introduce a separate prediction branch. Figure~\ref{fig:dcdn-operator} summarizes the resulting block.

\paragraph{Practical reciprocal.}
In the implemented models, the reciprocal branch is applied only to denominator-eligible coordinates and uses the stabilized reciprocal
$\rho_{\varepsilon}(x)=\operatorname{sgn}(x)/\max(|x|,\varepsilon)$ with $\varepsilon=10^{-2}$, preventing near-zero inputs from producing unbounded activations. Coordinates outside the reciprocal mask contribute zero through this branch.

\subsection{Laurent interaction support}
We now characterize the interaction family generated by stacking the ideal operator. Assume $x_i\neq 0$ for every input coordinate. For an exponent vector $\alpha\in\mathbb{Z}^d$, write
\begin{equation}
    \label{eq:alpha}
  x^\alpha=\prod_{i=1}^{d}x_i^{\alpha_i},
  \qquad
  \|\alpha\|_1=\sum_{i=1}^{d}|\alpha_i|,
\end{equation}
and define
\begin{equation}
\begin{aligned}
\mathcal{A}_K={}&\left\{\alpha\in\mathbb{Z}^d:
\|\alpha\|_1\le K+1,\ \max_i\alpha_i>0\right\}\\
&\cup\left\{\alpha\in\mathbb{Z}^d:
\alpha_i\le 0\ \forall i,\ \|\alpha\|_1\le K\right\}.
\end{aligned}
\label{eq:admissible-support}
\end{equation}

\begin{theorem}[Laurent support of a $K$-layer DCDN]\label{thm:laurent-support}
Consider a stack of $K\ge 1$ DCDN layers of the form in Eq.~\eqref{eq:dcdn-layer}, including the bias in the reciprocal branch, and use the exact element-wise reciprocal $1/x$. Every coordinate of $x^{(K)}$ is a Laurent polynomial whose monomial exponents belong to $\mathcal{A}_K$. Conversely, for every $\alpha\in\mathcal{A}_K$, there exists a choice of DCDN parameters for which the monomial $x^\alpha$ occurs with a nonzero coefficient in at least one coordinate of $x^{(K)}$.

If signed total degree is defined as $\deg_{\Sigma}(x^\alpha)=\sum_i\alpha_i$, then every monomial produced by the stack satisfies
\begin{equation}
  -K\le \deg_{\Sigma}(x^\alpha)\le K+1,
\end{equation}
and both bounds are attainable.
\end{theorem}

\begin{figure}[htbp]
    \centering
    \includegraphics[width=0.9\linewidth]{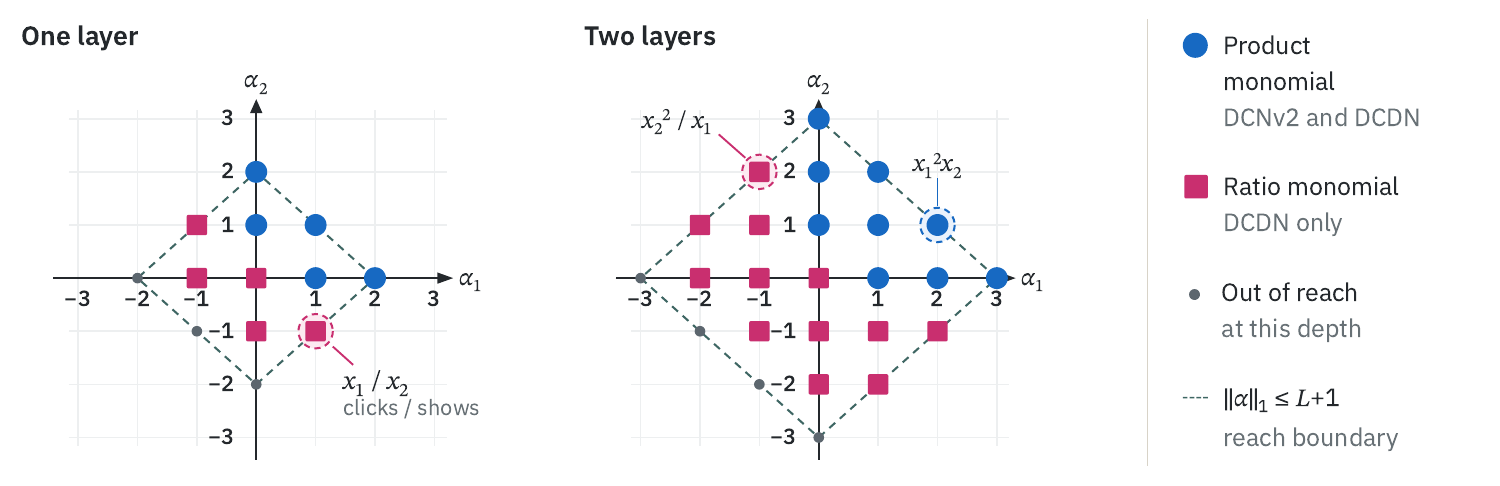}
    \caption{
        Reachable exponent lattice for two input features.
        Multiplicative and reciprocal crossings shift exponents in
        positive and negative coordinate directions, respectively. The notation of $\alpha$ is described in Equation~\eqref{eq:alpha}.
    }
    \Description{
        Two-dimensional integer exponent lattice illustrating the
        Laurent monomials reachable by stacked DCDN layers.
    }
    \label{fig:laurent-lattice}
\end{figure}

The lattice view makes the result intuitive. A standard cross increments an exponent, a reciprocal cross decrements it, and the residual connection leaves it unchanged. A single layer can therefore create terms such as $x_a/x_b$, while deeper stacks produce mixed positive- and negative-order interactions. Figure~\ref{fig:laurent-lattice} illustrates this geometry for two input coordinates; the proof of Theorem~\ref{thm:laurent-support} appears in Appendix~\ref{app:laurent-proof}.

All experiments use a single-layer form ($K=1$), which we refer to as the DCDN block.
With an exact reciprocal,
this layer generates every pairwise product $x_lx_k$ and, for masked denominators $x_k$, every ratio $x_l/x_k$ and reciprocal $1/x_k$.

\FloatBarrier
\section{Experimental Setup}
\subsection{Research questions}
The experiments address four research questions:

\begin{itemize}[leftmargin=*,nosep]
  \item \textbf{RQ1 (quotient learning):} Can DCDN learn ratio-like functions directly from their operands?
  \item \textbf{RQ2 (quality and efficiency):} How does DCDN compare with other feature-interaction models in ranking quality and measured inference latency?
  \item \textbf{RQ3 (capacity):} Are the observed differences explained simply by model capacity, or do they persist against a capacity-matched DCNv2 control?
  \item \textbf{RQ4 (production):} Do the quality and serving-efficiency advantages of DCDN transfer to production?
\end{itemize}

\subsection{Public tasks and feature views}
\label{sec:ptfv42}
We evaluate the architecture on Yambda-50M~\cite{ploshkin2025yambda}, KuaiRec~\cite{gao2022kuairec}, KuaiRand~\cite{gao2022kuairand}, and VK-LSVD~\cite{poslavsky2026vklsvd}. Table~\ref{tab:public-task-contracts} summarizes the target and evaluation protocol for each dataset;
it lists the dimensionality of both feature views and the number of base features that act as ratio denominators and are covered by the reciprocal mask.

\paragraph{Target choice.}
For Yambda-50M and VK-LSVD, we follow the binary targets used in the benchmark experiments of the dataset papers: \emph{Listen+}, defined by listening to at least $50\%$ of a track, and watch time above $10$ seconds, respectively. For KuaiRec, we use the binary construction recommended by the dataset authors for CTR-style evaluation, $\texttt{watch\_ratio}>2$. KuaiRand provides many feedback signals rather than a single canonical target; we use continuous watch time because it is the closest public analogue of the main consumption objective in our production setting. Yambda, KuaiRec, and VK-LSVD are evaluated with AUC and nDCG@10. For KuaiRand, XAUC is the primary ranking metric. Because the released KuaiRand log does not contain session identifiers, we reconstruct them using the common 30-minute inactivity heuristic: interactions are ordered chronologically for each user, and a new session starts whenever the gap from the preceding interaction exceeds 30 minutes~\cite{wu2017returning}. We report nDCG@10 over these reconstructed sessions and use it in RQ3 as a common top-ranking metric across all four datasets.

For each public dataset, we split interactions chronologically into training, validation, and test sets. Interaction-derived counter features are computed over historical lookback windows ending strictly before the timestamp of the interaction being scored. Thus, neither the target interaction nor future interactions contribute to its features, preventing temporal information leakage.

\paragraph{Ranking metrics.}
We use XAUC~\cite{zhan2022deconfounding}, defined as
the proportion of correctly ordered pairs
with distinct ground-truth targets:
\begin{equation}
  \operatorname{XAUC}
  = \frac{1}{|\mathcal{P}|}
    \sum_{(i,j)\in\mathcal{P}}
    \mathbf{1}\!\left[(\hat{y}_i-\hat{y}_j)(y_i-y_j)>0\right],
  \label{eq:xauc}
\end{equation}
where $\mathcal{P}$ contains comparable pairs ($y_i\neq y_j$).
For production XAUC, pairs are further restricted to the same
session. We compute nDCG with logarithmic position discount:
\begin{equation}
  \operatorname{nDCG}@10 =
  \frac{\sum_{i=1}^{m} g(y_{\pi(i)})/\log_2(i+1)}
       {\sum_{i=1}^{m} g(y_{\pi^*(i)})/\log_2(i+1)},
  \label{eq:ndcg}
\end{equation}
where $m=\min(10,n)$, $n$ is the evaluation group size,
$\pi$ and $\pi^*$ are the predicted and ideal rankings,
respectively. We use linear gain $g(y)=y$ for both binary
and continuous watch-time targets.

\paragraph{Loss normalization.}
For the binary tasks we use normalized entropy (NE)~\cite{he2014practical}, i.e., normalized cross-entropy in the terminology commonly used for industrial CTR models: binary cross-entropy divided by the entropy of the constant base-rate predictor, $H(\bar y)$. For KuaiRand we use normalized MSE (NMSE), defined as MSE divided by the target variance. Both normalization constants are precomputed from the training labels and then fixed.

\begin{table}[htbp]
\caption{Public tasks and feature views. $F_{\mathrm{reduced}}$ retains all base operands; $F_{\mathrm{full}}$ adds engineered ratio features.}
\label{tab:public-task-contracts}
\centering\small
\setlength{\tabcolsep}{3pt}
\begin{tabularx}{\linewidth}{@{}lYlrrrl@{}}
\toprule
Dataset & Target & \shortstack[l]{Main\\evaluation} & \shortstack{$\dim$\\$F_{\mathrm{reduced}}$} & \shortstack{$\dim$\\$F_{\mathrm{full}}$} & Denominators & Loss\\
\midrule
Yambda & played ratio $\ge 0.5$ (Listen+) & AUC, nDCG@10 & 86 & 134 & 12 & NE\\
KuaiRec & $\texttt{watch\_ratio}>2$ & AUC, nDCG@10 & 50 & 96 & 5 & NE\\
KuaiRand & watch time (s) & XAUC, nDCG@10 & 132 & 244 & 24 & NMSE\\
VK-LSVD & $\texttt{time\_spent}>10$ s & AUC, nDCG@10 & 99 & 216 & 36 & NE\\
\bottomrule
\end{tabularx}
\end{table}

\subsection{Models, training, and runtime measurement}
\label{sec:setup-models}
The comparison set contains MLP, DeepFM, DCN, AutoInt+, DCNv2, AFN+, EulerNet, Wukong, and DCDN. The main RQ2 quality comparison uses $F_{\mathrm{full}}$ for every model, so differences cannot be attributed to one architecture receiving explicit ratio features that another does not. The focused latency experiment uses $F_{\mathrm{reduced}}$ to compare model-side execution once the materialized ratios are removed.

For RQ3, we construct a \emph{capacity-matched DCNv2} control on the same reduced input $F_{\mathrm{reduced}}$ used by DCDN. We increase the depth of the DCNv2 cross stack until its trainable non-embedding parameter count is approximately matched to DCDN, while leaving the feature set unchanged. This control asks whether the behavior of DCDN can be reproduced by allocating a comparable parameter budget to the original architecture on the same input.

For each public dataset--model arm, the learning rate is selected once on a dedicated tuning seed from $\{10^{-4},3\!\times\!10^{-4},10^{-3},3\!\times\!10^{-3}\}$ using validation loss, and is then fixed for reported runs. We use AdamW, batch size $4096$, and at most $3000$ optimization steps. Validation is evaluated every $150$ steps, with early stopping after five consecutive validation checks without improvement. Categorical fields use learnable $16$-dimensional embeddings. Each final configuration is trained with eight seeds, and all reported public-data quality values are means across those runs. Across the four public datasets, the DCDN--DCNv2 improvements in the ranking metrics were statistically significant in seed-level comparisons ($p<0.05$).

Training and controlled inference measurements use NVIDIA A100 GPUs. For the model-only latency benchmark, we time eager execution at a fixed batch size of $8192$ under the same benchmark harness for every architecture and report p50 latency. This benchmark isolates model execution and is intentionally separate from the production measurements in Section~\ref{sec:prod-serving}, which include the deployed serving stack.

\FloatBarrier
\section{Offline Evaluation}
\subsection{RQ1: learning a quotient from operands}
Theorem~\ref{thm:laurent-support} shows that quotient-like terms are representable by DCDN. RQ1 asks the complementary optimization question: if the quotient itself is not provided as an input, does the reciprocal parameterization make that dependency easy to recover from data?

We evaluate methods on a synthetic task with eight positive operands $[n_1,d_1,n_2,d_2,d_3,z_1,z_2,z_3]$ and target $y=n_1/d_1$; $z_1,z_2,z_3$ are distractors and no model receives the target quotient as a feature. DCDN's reciprocal mask covers $d_1,d_2,d_3$. Operands are drawn log-uniformly from $[0.1,1000]$. We use this multi-scale distribution because production counters span orders of magnitude.

We evaluate four regimes: a \emph{reference range} with operands in $[1,10]$; \emph{common rescaling}, with all reference operands multiplied by $10$ while the quotient is unchanged; \emph{large denominators} in $[20,60]$; and \emph{small denominators} in $[0.2,0.6]$. The last two probe small- and large-valued quotients without changing the target definition.

Table~\ref{tab:rq1-summary} shows the result. DCDN is best on the
reference, large-denominator, and small-denominator regimes and third
on common rescaling. Its normalized MSE stays between
$2.81\times10^{-7}$ and $1.21\times10^{-6}$. Relative to DCNv2, the
error is lower by roughly $129\times$ on the reference range,
$1{,}620\times$ under common rescaling, $729\times$ with large
denominators, and $690\times$ with small denominators.

\begin{table}[htbp]
\caption{RQ1 normalized MSE (lower is better) after log-uniform training on $[0.1,1000]$. Reference uses operands in $[1,10]$; common rescaling multiplies all reference operands by $10$; the last two columns vary the denominator range.}
\label{tab:rq1-summary}
\centering\small
\setlength{\tabcolsep}{5pt}
\begin{tabular}{lrrrr}
\toprule
Model & \shortstack{Reference\\range} & \shortstack{Common\\rescaling} & \shortstack{Large\\denominator} & \shortstack{Small\\denominator}\\
\midrule
MLP & $6.18\times10^{-6}$ & $1.77\times10^{-5}$ & $9.36\times10^{-6}$ & $4.51\times10^{-3}$\\
DeepFM & $2.13\times10^{-6}$ & $2.81\times10^{-6}$ & $2.34\times10^{-6}$ & $3.58\times10^{-5}$\\
DCN & $2.91\times10^{-5}$ & $7.03\times10^{-4}$ & $2.83\times10^{-5}$ & $6.21\times10^{-4}$\\
AutoInt+ & $1.99\times10^{-6}$ & {\boldmath$8.01\times10^{-7}$} & $4.21\times10^{-7}$ & $3.13\times10^{-5}$\\
DCNv2 & $5.47\times10^{-5}$ & $1.96\times10^{-3}$ & $2.05\times10^{-4}$ & $6.03\times10^{-4}$\\
DCNv2-capacity-matched & $2.37\times10^{-5}$ & $1.19\times10^{-3}$ & $2.80\times10^{-4}$ & $2.50\times10^{-4}$\\
AFN+ & $8.55\times10^{-6}$ & $1.82\times10^{-5}$ & $1.06\times10^{-5}$ & $8.08\times10^{-4}$\\
EulerNet & $1.25\times10^{-6}$ & $9.79\times10^{-7}$ & $9.31\times10^{-7}$ & $8.19\times10^{-5}$\\
Wukong & $8.66\times10^{-6}$ & $5.72\times10^{-6}$ & $5.06\times10^{-6}$ & $2.32\times10^{-5}$\\
\textbf{DCDN} & {\boldmath$4.25\times10^{-7}$} & $1.21\times10^{-6}$ & {\boldmath $2.81\times10^{-7}$} & {\boldmath$8.74\times10^{-7}$}\\
\bottomrule
\end{tabular}
\end{table}

The common-rescaling is a sanity check as the target
is invariant to scaling both operands equally. DCDN remains accurate, although AutoInt+ and
EulerNet achieve slightly lower error. All reported regimes lie
within the log-uniform training support, so we do not interpret this
experiment as unrestricted arithmetic extrapolation.

\paragraph{Answer to RQ1.}
DCDN learns quotient-like functions directly from their operands with
high accuracy over the studied multi-scale range. It is best in
three of the four evaluation regimes and reduces DCNv2's error by
two to three orders of magnitude, demonstrating that reciprocal
crossing provides a strong inductive bias for learning ratio-like
dependencies beyond mere representability.

\subsection{RQ2: ranking quality}
We next evaluate the same architecture on recommendation data. All models in Table~\ref{tab:public-quality} receive $F_{\mathrm{full}}$, so this comparison isolates the architecture while holding the available input information fixed.

\begin{table}[htbp]
  \caption{Full-view public quality. AUC/XAUC is the primary metric; nDCG@10 is also reported for all four datasets. KuaiRand nDCG@10 uses sessions reconstructed with the 30-minute inactivity threshold described in Section~\ref{sec:ptfv42}. Higher is better.}
  \label{tab:public-quality}
  \centering
  \small
  \setlength{\tabcolsep}{2.7pt}
  \begin{tabular}{lcccccccc}
    \toprule
    & \multicolumn{2}{c}{Yambda} & \multicolumn{2}{c}{KuaiRec} & \multicolumn{2}{c}{KuaiRand} & \multicolumn{2}{c}{VK-LSVD} \\
    \cmidrule(lr){2-3}\cmidrule(lr){4-5}\cmidrule(lr){6-7}\cmidrule(lr){8-9}
    Model & AUC & nDCG@10 & AUC & nDCG@10 & XAUC & nDCG@10 & AUC & nDCG@10 \\
    \midrule
    MLP & 0.82110 & \textbf{0.7628} & 0.80067 & 0.4251 & 0.54830 & 0.6779 & 0.72203 & 0.8512 \\
    DeepFM & 0.82139 & 0.7602 & 0.80724 & 0.4357 & 0.54274 & 0.6679 & 0.72163 & 0.8526 \\
    DCN & 0.82010 & 0.7588 & 0.79992 & 0.4171 & 0.51457 & 0.6673 & 0.72422 & 0.8522 \\
    AutoInt+ & 0.82143 & 0.7623 & 0.80725 & 0.4393 & \textbf{0.55748} & 0.6797 & 0.72371 & 0.8528 \\
    DCNv2 & 0.81989 & 0.7560 & 0.80482 & 0.4260 & 0.51854 & 0.6718 & 0.71303 & 0.8505 \\
    AFN+ & 0.82158 & 0.7617 & 0.80533 & 0.4377 & 0.55552 & 0.6776 & 0.72427 & 0.8544 \\
    EulerNet & \textbf{0.82190} & 0.7614 & 0.80743 & 0.4408 & 0.53552 & \textbf{0.6843} & 0.72343 & 0.8515 \\
    Wukong & 0.82172 & 0.7625 & 0.80619 & 0.4343 & 0.53745 & 0.6748 & 0.71895 & 0.8504 \\
    \textbf{DCDN} & 0.82065 & 0.7603 & \textbf{0.80774} & \textbf{0.4457} & 0.52935 & 0.6771 & \textbf{0.72642} & \textbf{0.8545} \\
    \bottomrule
  \end{tabular}
\end{table}

With the same $F_{\mathrm{full}}$ inputs, DCDN improves both the primary ranking metric and nDCG@10 over DCNv2 on all four datasets (Table~\ref{tab:public-quality}). This consistent advantage over its direct architectural parent motivates the broader comparison with other models.

The broader comparison is more informative than a simple win count. DCDN is the best model on both reported metrics for KuaiRec and VK-LSVD. On Yambda, the strongest models are tightly clustered and EulerNet has the highest AUC; DCDN remains above DCNv2 but is not top-ranked. KuaiRand is the clearest counterexample to universal superiority: AutoInt+ has the highest XAUC, while EulerNet has the highest reconstructed-session nDCG@10. Relative to the generalized-order baselines that are closest in spirit, DCDN outperforms AFN+ and EulerNet on KuaiRec and VK-LSVD, while they are stronger on Yambda and KuaiRand. We therefore view DCDN as a consistently stronger replacement for DCNv2 and a competitive specialized alternative to more general interaction models, rather than as a universal accuracy winner.

\subsection{RQ2: inference efficiency}

We measure model-only inference latency with the protocol of Section~\ref{sec:setup-models} (eager execution on $F_{\mathrm{reduced}}$, batch size $8192$, p50), which excludes feature retrieval, request routing, and production batching.

We compare DCDN with the reduced-input DCNv2, the capacity-matched
DCNv2 control, AFN+, and EulerNet. The latter two are
relevant because, as discussed in Section~2, both can represent negative
or non-integer interaction orders using general parameterizations.
Table~\ref{tab:serving-capacity} shows a clear cost separation. DCDN
remains in the same sub-millisecond regime as DCNv2 on all four datasets: it is within $2.1\%$ of DCNv2 on Yambda and KuaiRec,
and is faster on KuaiRand and VK-LSVD. By contrast, AFN+ requires
$5.9\times$--$7.7\times$ the DCDN latency and EulerNet
$7.1\times$--$8.8\times$.

\paragraph{Answer to RQ2.}
DCDN provides a favorable quality--efficiency trade-off. It consistently
improves ranking quality over its direct DCNv2 parent while remaining in
approximately the same latency regime, and it is substantially cheaper
to serve than the more general AFN+ and EulerNet alternatives. The
result is therefore not universal accuracy dominance, but competitive
ranking quality at a much lighter inference cost.

\subsection{RQ3: capacity-matched comparison}
DCDN contains an additional reciprocal projection in each cross block, so one possible explanation for its behavior is simply that it changes the capacity of the interaction stack. We therefore compare it with the capacity-matched DCNv2 defined in Section~\ref{sec:setup-models}. Both models receive exactly the same $F_{\mathrm{reduced}}$ inputs; the control uses a deeper standard DCNv2 cross stack chosen to provide a comparable parameter budget. To make this comparison uniform across all four datasets, Table~\ref{tab:serving-capacity} uses nDCG@10 as the ranking-quality metric throughout.

\begin{table}[htbp]
\caption{Focused reduced-view serving and capacity comparison. Latency is eager-mode p50 in milliseconds at batch size $8192$. All quality columns use nDCG@10; KuaiRand nDCG@10 is computed on sessions reconstructed with a 30-minute inactivity threshold. $\Delta$ is DCDN-reduced minus capacity-matched DCNv2.
Suffix -r: $F_{\mathrm{reduced}}$; DCNv2-cm: capacity-matched DCNv2.}
\label{tab:serving-capacity}
\centering\small
\setlength{\tabcolsep}{2.4pt}
\begin{tabular}{lrrrrrrrrr}
\toprule
& \multicolumn{5}{c}{p50 latency (ms)} & \multicolumn{3}{c}{nDCG@10} & \\
\cmidrule(lr){2-6}\cmidrule(lr){7-9}
Dataset & DCDN-r & DCNv2-r & AFN+-r & EulerNet-r & DCNv2-cm & DCNv2-cm & DCDN-r & $\Delta$ & \shortstack{Speedup vs\\DCNv2-cm}\\
\midrule
Yambda & 0.622 & \textbf{0.614} & 4.331 & 5.196 & 0.718 & 0.7571 & \textbf{0.7590} & $+0.0019$ & $1.15\times$\\
KuaiRec & 0.635 & \textbf{0.622} & 3.752 & 4.539 & 0.731 & 0.4375 & \textbf{0.4441} & $+0.0066$ & $1.15\times$\\
KuaiRand & \textbf{0.713} & 0.816 & 5.512 & 6.294 & 1.045 & 0.6719 & \textbf{0.6765} & $+0.0046$ & $1.47\times$\\
VK-LSVD & \textbf{0.644} & 0.663 & 4.648 & 5.407 & 0.838 & 0.8521 & \textbf{0.8535} & $+0.0014$ & $1.30\times$\\
\bottomrule
\end{tabular}
\end{table}

DCDN has higher nDCG@10 on all four datasets and is
$1.15\times$--$1.47\times$ faster than the capacity-matched DCNv2.

\paragraph{Answer to RQ3.}
The observed operating point is not explained by parameter count alone.
With the same reduced input and a comparable capacity, DCDN
achieves higher nDCG@10 and lower p50 latency than the capacity-matched
DCNv2 on all four datasets. Simply deepening the original DCNv2 cross
stack therefore does not reproduce the same top-ranking
quality--latency trade-off.

\FloatBarrier
\section{Industrial Application}
\label{sec:industrial}
The public offline experiments isolate the interaction block under controlled conditions.
RQ4 asks whether the same quality--efficiency trade-off remains useful in the production system. We therefore evaluate a local intervention in the existing ranking pipeline rather than redesigning the surrounding recommender.

\subsection{Production ranking architecture}
\label{sec:prod-architecture}
The production system follows a candidate-generation--ranking architecture. The final ranker consumes heterogeneous inputs: counter-like user, item, and context statistics; predictions produced by upstream models; and sparse categorical features. The control feature pipeline also materializes a sizeable family of manually engineered ratio features derived from these signals, including normalized counters and ratios involving upstream-model scores. These features are useful to the existing ranker but increase both feature-pipeline complexity and the width of the interaction stage.

Continuous and count-like inputs are encoded with piecewise-linear encoding~\cite{gorishniy2022embeddings}, while sparse categorical features are represented through embedding lookup. The deployed treatment keeps the underlying operands and upstream-model predictions but removes the targeted engineered ratios. The encoded representation is then passed through a single DCDN layer, followed by \hiformer~\cite{gui2023hiformer}. The two interaction stages play different roles: DCDN provides a lightweight structured bias toward product/ratio relations, whereas \hiformer's heterogeneous self-attention captures more general cross-feature dependencies.

\begin{figure}[htbp]
  \centering
  \includegraphics[width=0.98\linewidth]{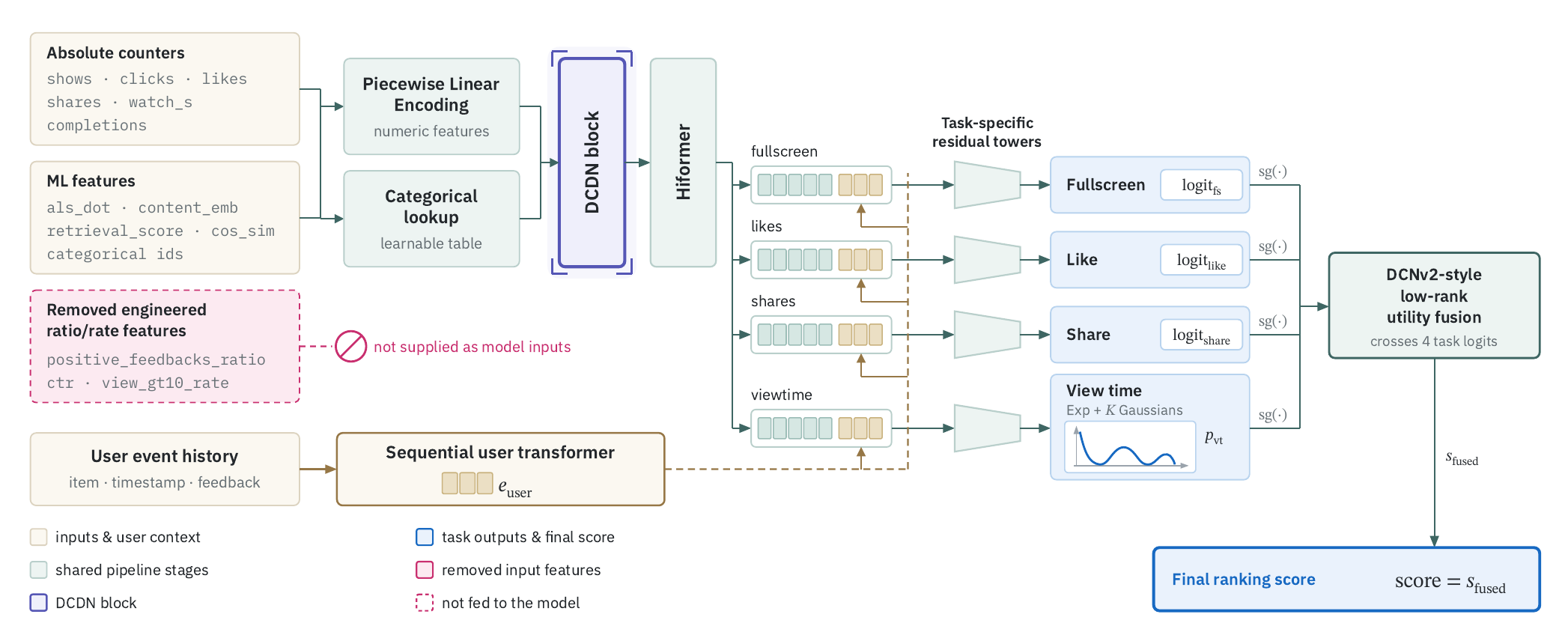}
  \caption{Production ranking architecture used by the treatment. Relative to control, the treatment removes the highlighted engineered ratio/rate inputs and replaces the DCNv2 interaction block with a single DCDN layer. The downstream heterogeneous-interaction, user-model, task-head, and scoring components are kept fixed.}
  \Description{Production ranking diagram. Counter-like and upstream-model features are encoded by piecewise-linear and categorical embedding stages, while selected engineered ratio/rate features are removed from the treatment input. A DCDN block is followed by Hiformer, task-specific residual towers, and a sequential-user representation. Task outputs are combined by the same downstream scoring logic used in the control.}
  \label{fig:industrial-ranker}
\end{figure}

The resulting representation feeds task-specific residual towers for the production objectives, including watch-time distribution modeling~\cite{zhao2025multi,gulevskaia2026hierarchical}, likes, shares, and clicks to open fullscreen. A sequential user transformer supplies a user-history embedding to the task predictors. The downstream calibration and utility-fusion logic is identical in control and treatment; these components are therefore not part of the DCDN intervention. Figure~\ref{fig:industrial-ranker} summarizes the complete production architecture and highlights the feature path removed relative to control.

\subsection{Training and offline deployment gate}
\label{sec:prod-training}
Each production training run uses six days of logged interactions followed by one day of validation. Because the ranker optimizes several engagement objectives, we use a shared multi-head model rather than independent task-specific rankers. A multi-head architecture allows us to share the expensive representation path across objectives while avoiding multiple forward passes at runtime.

The control and treatment differ only in the interaction
block and the associated engineered ratio-feature inputs. This choice is inherited from the design of the predecessor model rather than from a DCDN-specific depth search. In the architecture study for that ranker, one explicit cross layer followed by \hiformer provided sufficient expressiveness for the downstream objectives, and DCNv2 outperformed DCN and Wukong in the resulting quality--latency comparison. We therefore keep this established operating point and change only the cross operator and the associated ratio-feature path.

Before online deployment, we compare four offline variants on production validation data using session-level XAUC.
The four variants are DCNv2 and DCDN under the full feature view $F_{\mathrm{full}}$, and the same two architectures under the reduced feature view $F_{\mathrm{reduced}}$. Among these comparisons, the deployed production pair is \textit{DCNv2 full} (control) versus \textit{DCDN reduced} (treatment); the full/full and reduced/reduced comparisons are auxiliary offline checks that separate feature-view and architecture effects.

\begin{figure}[htbp]
  \centering
  \includegraphics[width=0.62\linewidth] {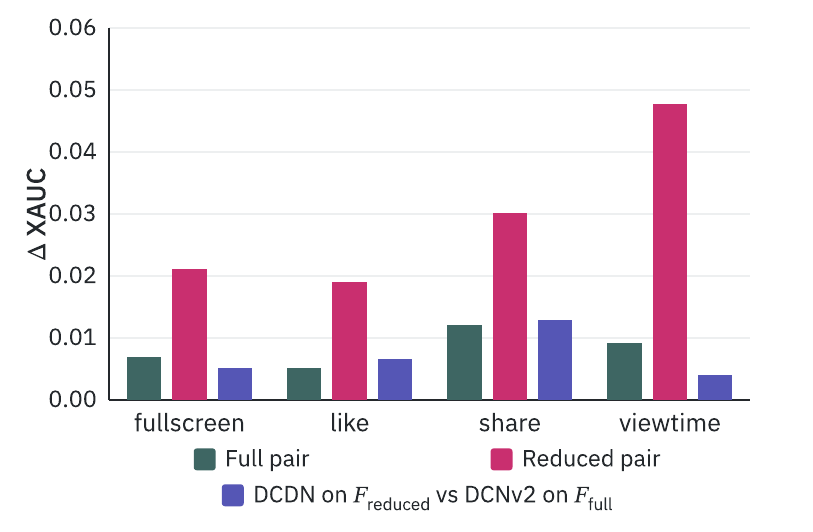}
  \caption{Production offline XAUC comparison across full and reduced feature views.
  For each objective, the first two bars compare DCDN with DCNv2 under the same feature view (full, reduced); the third bar is the deployed comparison, DCDN reduced minus DCNv2 full.}
  \Description{Grouped comparison chart for the production offline study. For each objective---fullscreen, like, share, and view time---the chart shows three deltas in XAUC: DCDN full minus DCNv2 full, DCDN reduced minus DCNv2 reduced, and DCDN reduced minus DCNv2 full. All deltas are positive, and the reduced-feature comparison gives the largest gains.}
  \label{fig:prod-offline-comparison}
\end{figure}

Table~\ref{tab:prod-offline} shows that DCDN outperforms DCNv2 on all four production objectives under both matched feature views. Crucially, the deployed treatment (DCDN reduced) also improves XAUC over the production control (DCNv2 full) on every objective, despite removing $40.1\%$ of input features. Figure~\ref{fig:prod-offline-comparison} illustrates these differences.

\begin{table}[htbp]
\caption{Production offline quality measured by session-level XAUC under full and reduced feature views. Higher is better. ``Full pair'' is DCDN full minus DCNv2 full; ``Reduced pair'' is DCDN reduced minus DCNv2 reduced; ``Production pair'' is DCDN reduced minus DCNv2 full, i.e., the deployed treatment--control comparison.}
\label{tab:prod-offline}
\centering\small
\setlength{\tabcolsep}{4pt}
\begin{tabular}{lccccccc}
\toprule
& \multicolumn{4}{c}{Absolute XAUC} & \multicolumn{3}{c}{$\Delta$ XAUC}\\
\cmidrule(lr){2-5}\cmidrule(lr){6-8}
Objective & \shortstack{DCNv2\\full} & \shortstack{DCDN\\full} & \shortstack{DCNv2\\reduced} & \shortstack{DCDN\\reduced} & \shortstack{Full\\pair} & \shortstack{Reduced\\pair} & \shortstack{Production\\pair}\\
\midrule
Fullscreen & 0.6238 & \textbf{0.6307} & 0.6077 & {0.6289} & $+0.0069$ & $+0.0212$ & $+0.0051$\\
Like & 0.7325 & 0.7377 & 0.7200 & \textbf{0.7391} & $+0.0052$ & $+0.0191$ & $+0.0066$\\
Share & 0.5805 & 0.5926 & 0.5632 & \textbf{0.5934} & $+0.0121$ & $+0.0302$ & $+0.0129$\\
View time & 0.5589 & \textbf{0.5681} & 0.5152 & {0.5629} & $+0.0092$ & $+0.0477$ & $+0.0040$\\
\bottomrule
\end{tabular}
\end{table}

\subsection{Online A/B test}
\label{sec:prod-ab}
We ran a two-week A/B experiment on production traffic. The control group used the existing ranker with a DCNv2 interaction block and the corresponding engineered ratio/rate inputs. The treatment used the same ranking pipeline with DCNv2 replaced by DCDN and the targeted engineered ratios removed. All downstream task heads and scoring logic were kept unchanged. Users were randomly assigned to control and treatment. Statistical significance was assessed using two-sided $t$-tests on CUPED-adjusted user-level metric aggregates, with pre-experiment observations used for variance reduction. All relative changes in Table~\ref{tab:ab-results} are statistically significant ($p<0.01$).

\begin{figure}[htbp]
  \centering
  \begin{minipage}{0.49\linewidth}\centering\includegraphics[width=\linewidth]{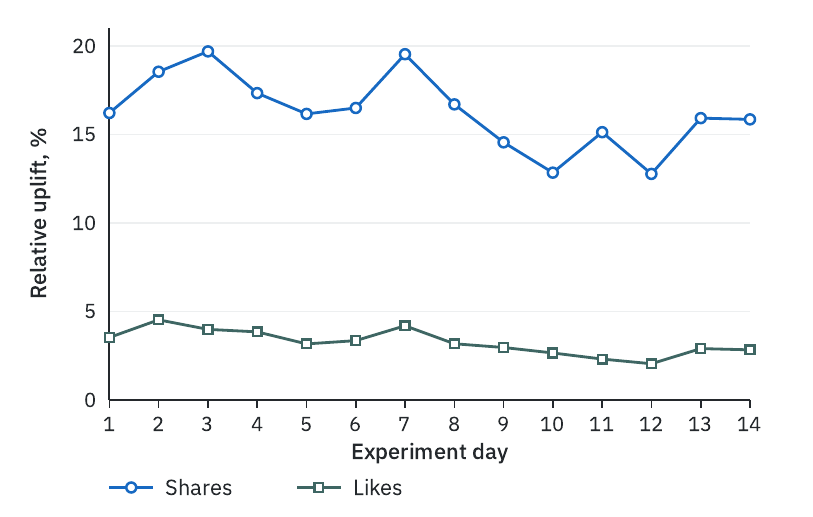}\end{minipage}\hfill
  \begin{minipage}{0.49\linewidth}\centering\includegraphics[width=\linewidth]{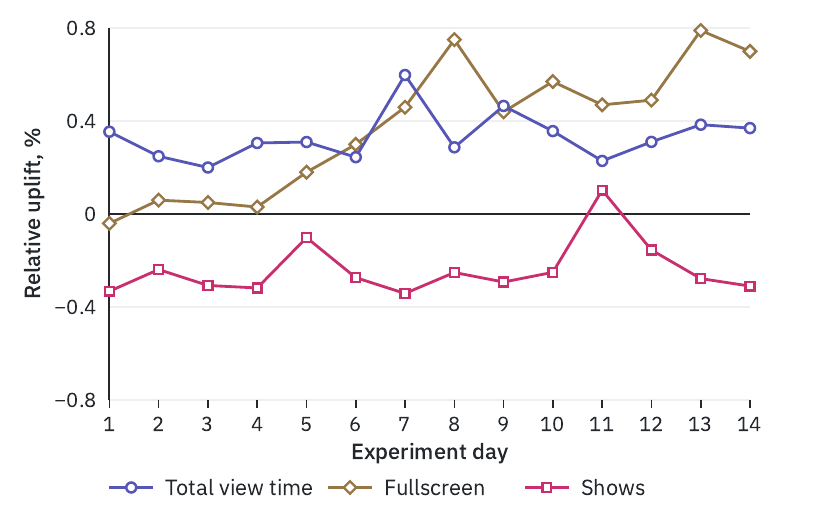}\end{minipage}
  \caption{
  Daily relative treatment changes over the 14-day A/B window. Left: shares and likes; right: total view time (TVT), fullscreen opens, and shows (different y-scales).}
  \Description{Daily A/B deltas from July 29 to August 11, 2026.}
  \label{fig:ab-daily}
\end{figure}

Table~\ref{tab:ab-results} reports aggregate relative changes over the experiment. Total View Time (TVT) increases by $0.33\%$, likes by $3.25\%$, shares by $16.26\%$, and clicks to open fullscreen by $0.38\%$, while the number of shows changes by only $-0.24\%$. The day-level view in Figure~\ref{fig:ab-daily} provides a complementary stability check. Likes, shares and TVT are positive on all 14 days, and clicks to open fullscreen on 13 days.

\begin{table}[htbp]
\caption{Production A/B relative changes over the 14-day experiment.}
\label{tab:ab-results}
\centering\small
\begin{tabular}{lr}
\toprule
Metric & Change\\
\midrule
Total View Time (TVT) & $+0.33\%$\\
Likes & $+3.25\%$\\
Shares & $+16.26\%$\\
Shows & $-0.24\%$\\
Fullscreen & $+0.38\%$\\
\bottomrule
\end{tabular}
\end{table}

\FloatBarrier
\subsection{Production serving performance}
\label{sec:prod-serving}
The production ranker is served with Triton Inference Server using dynamic batch sizes from $1$ to $16{,}384$. The service operates at tens of thousands of ranking requests per second and supports a product with tens of millions of daily active users, making any per-request savings material at system scale. The serving implementation uses fused Triton kernels for feature encoding and an analogous fused implementation for DCDN. On NVIDIA RTX6000 Pro GPUs, the custom feature-encoding kernel is approximately $7\times$ faster than its eager counterpart, while the fused DCDN implementation is approximately $2\times$ faster than eager DCDN. We report these as component-level implementation results; they are distinct from the end-to-end treatment effect below.

At the deployed-package level, model-computation time decreases by
$42.4\%$ and ranking-service time by $10.5\%$
(Table~\ref{tab:prod-serving}); the smaller end-to-end gain reflects
other components of the serving stack.

\begin{table}[htbp]
\caption{Production efficiency changes.}
\label{tab:prod-serving}
\centering\small
\begin{tabular}{lr}
\toprule
Quantity & Change\\
\midrule
Number of input features & $-40.1\%$\\
Model computation time & $-42.4\%$\\
Ranking-service time & $-10.5\%$\\
\bottomrule
\end{tabular}
\end{table}

These are package-level gains: the treatment changes both the feature
path and interaction block, and realized latency depends on batching and
operator fusion. They therefore do not imply that reciprocal arithmetic
is intrinsically cheaper.

\paragraph{Answer to RQ4.}
The quality--efficiency advantage transfers to the deployed system.
Relative to the full-feature DCNv2 control, the reduced-feature DCDN
treatment improves offline XAUC on all production objectives and yields
statistically significant online gains in the main engagement metrics,
while removing $40.1\%$ of the input features. At the same time,
model-computation time decreases by $42.4\%$ and end-to-end
ranking-service time by $10.5\%$.

\FloatBarrier
\section{Practical Lessons}
\label{sec:practical-lessons}
\paragraph{Move recurring feature algebra into the model.}
DCDN targets manually engineered ratios whose operands remain
available. Encoding such recurring algebraic relations in the
interaction layer reduces feature-pipeline complexity and input
width without discarding the underlying signals. In production,
we remove $40.1\%$ of input features while preserving base
counters and upstream-model predictions.

\paragraph{Specialized and general interaction modules can be complementary.}
We found no need to make DCDN responsible for every cross-feature dependency in the production model. A single DCDN layer handles the product/ratio interactions targeted by the architecture, while \hiformer captures broader heterogeneous interactions downstream. This division of labor suggests that a specific algebraic inductive bias can be useful as one stage of a larger interaction stack rather than as a replacement for a general interaction learner.

\paragraph{Serving efficiency is an end-to-end property.}
Sharing the representation across objectives, dynamic batching,
and fused Triton kernels---not interaction-layer FLOPs alone---
determine serving performance. The $42.4\%$ reduction in
model-computation time yields a smaller $10.5\%$ reduction in
ranking-service time, highlighting the need for end-to-end
measurements alongside component microbenchmarks.

\FloatBarrier
\section{Discussion and Limitations}
Across public tasks, DCDN consistently outperforms DCNv2 but does not dominate the broader model family: AFN+ and EulerNet remain stronger on some datasets. Its main contribution is therefore a specialized inductive bias for product/ratio interactions that enables a favorable quality--efficiency trade-off, rather than universal accuracy superiority.

RQ1 helps explain why the specialized operator can work. Under a log-uniform training distribution spanning the multi-scale range of counter features, DCDN learns the target quotient to near-zero error across reference, rescaling, and denominator-shift regimes and is dramatically more accurate than DCNv2. This complements the Laurent-support theorem: the theory establishes representability, while the controlled experiment shows learnability in the numerical range of interest. The experiment does not establish arbitrary out-of-distribution arithmetic, since the reported regimes lie within the training support. Likewise, the clipped reciprocal used in practice differs from the exact Laurent operator outside its unclipped range.

Several limitations qualify these results. First, the controlled quotient experiment does not make DCDN uniformly best even in the setting it targets: under common rescaling, AutoInt+ and EulerNet attain lower NMSE, while DCDN is third-best. This suggests that the reciprocal bias makes quotient-like dependencies easy to learn, but does not guarantee the best optimization behavior for every transformation or architecture. Second, latency depends on kernels, hardware, and batching. Our eager-mode benchmark provides a controlled relative comparison, while the Triton result is the relevant systems evidence for our deployment; neither is a universal operator-complexity claim. Third, KuaiRand remains a clear counterexample to universal accuracy superiority:
six of the eight baselines, including MLP, achieve higher XAUC there.
Together with the different targets and evaluation protocols across the four public datasets, this heterogeneity argues against collapsing the results into a single aggregate score.
Fourth, the reciprocal mask is specified from known ratio denominators; selecting eligible coordinates automatically is left for future work.

\FloatBarrier
\section{Conclusion}
DCDN adds a reciprocal direction to the CrossNet recursion, giving the model a compact way to form normalized interactions from the same operands used by standard crosses. The proposed operator has an exact Laurent-polynomial characterization, and the experiments show that this parameterization makes quotient-like dependencies easy to learn over the multi-scale range of interest. Across the public tasks, DCDN consistently improves on DCNv2 and is competitive with more general interaction models, while remaining in a much lighter inference regime than AFN+ and EulerNet. In production, the same design reduces reliance on manually engineered ratio features, improves several engagement metrics, and lowers both model-computation and ranking-service time.

More broadly, DCDN illustrates how model structure can absorb
recurring algebraic feature transformations through a compact
inductive bias.

\FloatBarrier
\appendix
\FloatBarrier
\section{Proof of Theorem~\ref{thm:laurent-support}}\label{app:laurent-proof}
\begin{proof}
Let $e_j$ denote the $j$-th standard basis vector in
$\mathbb{Z}^d$. Consider a monomial $x^\alpha$ occurring in some
coordinate of $x^{(\ell)}$. From Eq.~\eqref{eq:dcdn-layer}, its
contribution to output coordinate $j$ can have only one of three
exponent vectors: $\alpha$, $\alpha+e_j$ and $\alpha-e_j$.
Indeed, these correspond respectively to the residual connection,
multiplication by $x_j$, and multiplication by $x_j^{-1}$. The two
bias terms additionally introduce the monomials $x_j$ and $x_j^{-1}$.
Thus one DCDN layer either preserves an exponent vector or changes
exactly one of its coordinates by $\pm 1$.

We first characterize which exponent vectors can occur. At depth zero,
coordinate $s$ contains only $x_s=x^{e_s}$. Hence every monomial
descending from the original input after $K$ layers must satisfy $\|\alpha-e_s\|_1\le K$
for some $s$. The reciprocal bias can also create $x_s^{-1}=x^{-e_s}$;
creating this term consumes one layer, so any monomial descending from
such a term satisfies $\|\alpha+e_s\|_1\le K-1$
for some $s$. Therefore the support of $x^{(K)}$ is contained in
\[
\mathcal S_K
=
\bigcup_s B_1(e_s,K)
\;\cup\;
\bigcup_s B_1(-e_s,K-1),
\]
where $B_1(c,r)=\{\alpha\in\mathbb Z^d:\|\alpha-c\|_1\le r\}$.

This set is exactly $\mathcal A_K$ from
Eq.~\eqref{eq:admissible-support}. If $\alpha$ has a positive coordinate,
choose $s$ with $\alpha_s>0$. Then $\|\alpha-e_s\|_1=\|\alpha\|_1-1$,
so $\alpha$ belongs to the first union exactly when
$\|\alpha\|_1\le K+1$. If all coordinates are nonpositive and
$\alpha\neq0$, choose $s$ with $\alpha_s<0$. Then
$
    \|\alpha+e_s\|_1=\|\alpha\|_1-1,
$
so $\alpha$ belongs to the second union whenever
$\|\alpha\|_1\le K$. The zero vector is reachable from
$x_s/x_s=1$ for $K\ge1$.
Thus $\mathcal A_K \subseteq \mathcal S_K$.
For the reverse inclusion,
$\|\alpha\|_1\le\|\alpha\mp e_s\|_1+1$ gives $\|\alpha\|_1\le K+1$ on the first union and $\|\alpha\|_1\le K$ on the second, while a nonpositive $\alpha$ in the first union has $\|\alpha\|_1=\|\alpha-e_s\|_1-1\le K-1$; hence $\mathcal S_K=\mathcal A_K$.

The signed-degree bound follows directly from the same one-step
description. Initially every coordinate has signed degree $1$.
A layer changes $\sum_i\alpha_i$ by at most one, while the reciprocal
bias can introduce degree $-1$ already in the first layer. Thus after
one layer the possible signed degrees lie in $[-1,2]$, and every
subsequent layer can extend either endpoint by at most one. After
$K$ layers, $-K\le \sum_i\alpha_i\le K+1.$

It remains to show attainability. Set every entry of
$W_\ell,\widetilde W_\ell,b_\ell,\widetilde b_\ell$ to one. All
monomial coefficients generated by the recursion are then
nonnegative, so a term that is generated along any path cannot be
cancelled by another contribution. For any
$\alpha\in\mathcal A_K$ having a positive coordinate, choose
$s$ with $\alpha_s>0$. Since
$
    \|\alpha-e_s\|_1=\|\alpha\|_1-1\le K,
$
there exists a sequence of at most $K$ unit moves
$\pm e_j$ from $e_s$ to $\alpha$. Each such move is realized by the
corresponding multiplicative or reciprocal branch, while residual
connections fill any unused layers. If all coordinates of $\alpha$
are nonpositive and $\alpha\neq0$, first use the reciprocal bias to
create $-e_s$ for some $\alpha_s<0$; the remaining path has length $\|\alpha+e_s\|_1=\|\alpha\|_1-1\le K-1.$
Finally, $\alpha=0$ is produced by $x_s/x_s$. Hence every
$\alpha\in\mathcal A_K$ occurs with a nonzero coefficient.

The degree bounds are tight: repeatedly taking the multiplicative
branch produces $x_j^{K+1}$, while the reciprocal bias followed by
$K-1$ reciprocal steps produces $x_j^{-K}$.
\end{proof}

\FloatBarrier
\section*{Ethical Considerations}
This work combines public recommendation benchmarks with a production A/B experiment. Yambda-50M and VK-LSVD are released under the Apache License 2.0, while KuaiRec and KuaiRand are released under CC BY-SA 4.0. We use the released Yambda-50M subset, the KuaiRec Big Matrix, KuaiRand, and the public VK-LSVD interaction release.
We use KuaiRand-Pure, the release variant restricted to the $7{,}583$ videos in its candidate pool,
with user and video identifiers re-indexed; VK-LSVD publishes categorical identifiers as anonymized stable integers without a reverse mapping. Beyond the task definitions and feature transformations described in Section~4.2, we do not attempt to recover identities or enrich these datasets with external user-level information.

Production data are processed only inside access-controlled company infrastructure and in accordance with the company's data-protection and anonymization policies. The production model analyzed in this work does not use sensitive or protected user attributes; its inputs are interaction-derived, contextual, and model-derived signals of the type described in Section~\ref{sec:prod-architecture}. Experimental records use de-identified internal identifiers, and the authors have no access to mappings from those identifiers to real-world identities or to any mechanism for de-anonymizing users. Only aggregate treatment effects and coarse service-scale descriptors are reported; no user-level logs or exact traffic counts are disclosed.

The experiment was reviewed and conducted under the company's standard experimentation and data-governance process. The paper reports only aggregate treatment--control effects and does not disclose user-level observations or internal experimentation thresholds. No additional sensitive information was collected or used for the purpose of this research.

Recommendation changes can affect exposure and the balance among multiple user objectives even when aggregate engagement improves. We therefore report shows together with engagement outcomes and keep the downstream multi-objective scoring logic fixed between control and treatment. We do not perform a group-wise fairness analysis, so the reported aggregate gains do not establish fairness across demographic or creator groups. Likewise, improvements in watch time, likes, shares, or fullscreen opens should not be interpreted as a general user-welfare guarantee. These limitations, together with the finite two-week experiment horizon, bound the broader conclusions that can be drawn from the deployment.

\FloatBarrier
\bibliographystyle{IEEEtran}
\bibliography{references}

\end{document}